\documentstyle[epsfig]{europhys}
\def\openone{\leavevmode\hbox{\small1\normalsize\kern-.33em1}}%

\def\And{{\rm and\ }}

\def\drm{{\rm d}}

\def\dif#1#2{\frac{\drm#1}{\drm#2}}

\def\order{\mathop{\rm O}\nolimits}
\def\tr{\mathop{\rm tr}\nolimits}

\newif\ifboo \boofalse

\makeatletter
\def\lesssim{\mathrel{\mathpalette\vereq<}}
\def\vereq#1#2{\lower3pt\vbox{\baselineskip1.5pt \lineskip1.5pt
\ialign{$\m@th#1\hfill##\hfil$\crcr#2\crcr\sim\crcr}}}
\makeatother
\begin{document}
\euro{}{}{}{}
\Date{}
\shorttitle{K. FRAHM {\em et al.}\/: LARGE PETERMANN FACTOR IN CHAOTIC CAVITIES}

\title{Large Petermann factor in chaotic cavities with many 
scattering channels}
\author{K. Frahm\inst{1}, H. Schomerus\inst{2}, M. Patra\inst{2}
\And C. W. J. Beenakker\inst{2}}

\institute{
     \inst{1} Laboratoire de Physique Quantique, UMR 5626 du CNRS,
		 Universit\'e{} Paul Sabatier, F-31062 Toulouse Cedex 4, France
		 \\
     \inst{2} Instituut-Lorentz, Universiteit Leiden, P.O. Box 9506,
		 2300 RA Leiden, The Netherlands
}

\rec{}{} 

\pacs{
\Pacs{42}{50Lc}{Quantum fluctuations, quantum noise, and quantum jumps}
\Pacs{42}{50Ar}{Photon statistics and coherence theory}
\Pacs{42}{60Da}{Resonators, cavities, amplifiers, arrays, and rings}
}
\maketitle

\begin{abstract}

The quantum-limited linewidth of a laser cavity is
enhanced above the Schawlow-Townes value by the Petermann factor $K$, 
due to the non-orthogonality of the cavity modes.
The average 
Petermann factor $\langle K\rangle$
in an ensemble of cavities with chaotic scattering
and broken time-reversal symmetry
is calculated non-perturbatively using
random-matrix theory and the supersymmetry 
technique, as a function of the decay rate $\Gamma$ of the lasing mode and the 
number of scattering channels $N$.
We find for $N\gg 1$ that for typical values of $\Gamma$ 
the average Petermann factor
$\langle K\rangle\propto \sqrt{N}\gg 1$
is parametrically larger than unity.
\end{abstract}

The study of resonant scattering goes back to the early work of Breit
and Wigner \cite{BW}, and was developed extensively in the context of
nuclear physics \cite{MW}.
The Breit-Wigner resonance is described by a frequency-dependent
scattering matrix
$S(\omega)$ with elements
\begin{equation}
\label{eq:s}
S_{nm}=\delta_{nm}+\sigma_n^{}\sigma_m'(\omega-\Omega+i\Gamma/2)^{-1}
\ ,
\end{equation}
where $\Omega$ and $\Gamma$ are, respectively, the center and the width
of the resonance, and $\sigma_n^{}$, $\sigma_m'$ are the complex coupling
constants of the resonance to the scattering channels $n$, $m$.
(In the presence of time-reversal symmetry the scattering matrix
is symmetric, hence $\sigma_n^{}=\sigma'_n$.) The coupling constants of a
Breit-Wigner resonance are related to its width by the sum rule
\cite{Buettiker} 
\begin{equation}
\label{eq:sumrule}
\sum_{n,m}|\sigma_n^{}\sigma_m'|^2=\Gamma^2
\ .
\end{equation}
The physics behind this sum rule is that narrow resonances are quasi-bound
states requiring a weak coupling to the continuum.

The Breit-Wigner theory holds in the limit $\Gamma\to 0$ only.
More generally, one may consider a pole $\Omega-i\Gamma/2$
somewhere in the lower half of the complex $\omega$-plane and ask for
the form of the scattering matrix near that pole. The separable form 
(\ref{eq:s}) remains valid, even if $\Gamma$ is not small.
The sum rule  (\ref{eq:sumrule}), however, should be replaced  by
\cite{patra:99}
\begin{equation}
\label{eq:sr2}
\sum_{n,m}|\sigma_n^{}\sigma_m'|^2=K\Gamma^2\ ,\quad K\geq 1\ .
\end{equation}
The total coupling strength of the resonance is larger than required by
its width by a factor $K$.
This factor has been studied theoretically and
experimentally in laser cavities \cite{petermann:79a,siegman:89,cheng:96a,eijkelenborg:96a},
where it governs the quantum-limited linewidth,
\begin{equation}
\delta\omega=K\frac{\Gamma^2}{2I}
\ ,
\label{eq:k}
\end{equation}
of the radiation from the lasing
mode (with output power $I$). The linewidth is larger than the
Schawlow-Townes value $\Gamma^2/2I$ \cite{schawlow:58a} by the factor $K$.
In this context $K$ is called the Petermann factor or excess noise
factor \cite{petermann:79a}.

The optical cavities considered in these studies typically have a simple
shape, that is characterized by integrable rather than chaotic wave
dynamics \cite{haake:91}. Recently, three of us investigated the
statistics of the Petermann factor in the presence of chaotic scattering
\cite{patra:99}. That work was restricted to the case of a single
scattering channel ($N=1$), meaning a scalar scattering matrix. If the
cavity is coupled to the outside world via an opening of area ${\cal A}$,
then a single scattering channel requires ${\cal A}  \lesssim
\lambda^2/2\pi$, with $\lambda$ the wavelength of the radiation. 
The factor $K$ is only slightly larger than $1$ for such an almost closed
cavity, and measuring the deviation from unity would be quite
problematic. For a larger $K$ one has to increase ${\cal A}$,
thereby increasing the number of scattering channels
$N\approx 2\pi {\cal A}/\lambda^2$.

The fundamental question addressed in this paper is how $K$ increases
with increasing $N$. We present for the case of broken time-reversal
symmetry a non-perturbative calculation of the average Petermann factor
$\langle K\rangle$ in an ensemble of chaotic cavities, valid for
arbitrary $N$ and $\Gamma$. A non-perturbative calculation is necessary
because 
the mode selected for laser action has a decay rate
$\Gamma$ of the order of the typical smallest rate
$\Gamma_0=TN\Delta/2\pi$, with $\Delta$ the mean spacing in frequency of
the resonances and $T$ the transmission probability of the channels
through the opening.
(The time $1/\Gamma_0$ is the mean dwell time
of photons in the cavity.) Existing methods of large-$N$ perturbation
theory \cite{janik:99a} require that $\Gamma>\Gamma_0$, and are therefore
not applicable to this problem. Our non-perturbative method is based on
the supersymmetry technique \cite{efetov:97,fyodorov:97a}, applied to
eigenvector correlations of non-Hermitian random matrices.
Our conclusion is that $\langle K  \rangle\propto \sqrt N\gg 1$ can become
parametrically large for $N\gg 1$. The square-root increase is the
outcome of a subtle balance between the smallness of $\Gamma$ and the
largeness of $N$, and could not have been anticipated from perturbative
arguments, which would imply a linear increase.

To formulate the problem within the framework of
random-matrix theory we proceed as in ref.~\cite{patra:99}.
The modes of a closed chaotic cavity, in the absence of
time-reversal symmetry, are eigenvalues of an $M\times M$ Hermitian
matrix $H$ with independent Gaussian distributed elements
\cite{haake:91,guhr:98}. 
The limit $M\to\infty$ at fixed spacing $\Delta$ of 
the modes is taken at the end of the calculation. 
The opening in the cavity is described by an $M\times N$ 
coupling matrix $W$.
Modes of the open
cavity are complex eigenvalues of the 
non-Hermitian matrix ${\cal H}=H - i\pi W W^\dagger$ which determines
the scattering matrix \cite{MW,guhr:98}
\begin{equation}
	S = 1 - 2 \pi i W^\dagger ( \omega - {\cal H} )^{-1} W \ .
	\label{Smatrix}
\end{equation}
The matrix ${\cal H}=U\,\mbox{diag}\,(z_1,z_2,\ldots,z_M)U^{-1}$
has complex eigenvalues $z_j$ and the matrix U contains the 
right eigenvectors of ${\cal H}$ in its columns. 
(The matrix $U^{-1}$ contains the left eigenvectors.) Because 
${\cal H}$ is not Hermitian, $U$ is not
unitary and the left or right eigenvectors are not orthogonal among themselves.
If the cavity is filled with a homogeneous 
amplifying medium (amplification rate $1/\tau_a$) one has to add a term 
$i/2\tau_a$ to the eigenvalues, shifting them upwards towards the real axis,
while the eigenvectors are unchanged.
The lasing mode is the eigenvalue $z_l=\Omega-i\Gamma/2$ closest to the real
axis, and the laser threshold is reached when the decay rate $\Gamma$
of this mode equals the amplification rate $1/\tau_a$.
Near the laser threshold it is only necessary to retain the contribution from 
the lasing mode.
Combining eqs.~(\ref{eq:s}), (\ref{eq:sr2}), and (\ref{Smatrix})
one finds that the Petermann factor is given by
\cite{patra:99,siegman:89}
\begin{equation}
\label{Petfak1}
K=(U^\dagger\, U)_{ll}\,(U^{-1}\,U^{-1\dagger})_{ll}
\ . 
\end{equation}
This formula shows that deviations of $K$ from unity result from the
non-unitarity
of $U$, hence from the non-orthogonality of the cavity modes \cite{siegman:89}.

We will calculate the average Petermann factor at a given value
of $\Gamma$ and $\Omega$.
Following refs.\ \cite{janik:99a,chalker:98a}, we write this conditional
average in the form
\begin{eqnarray}
\label{Petfak2}
\langle K\rangle&=&
\Bigl\langle\sum_j 
(U^\dagger\, U)_{jj}\,(U^{-1}\,U^{-1\dagger})_{jj}\ \delta(z-z_j)\Bigr\rangle
\Bigl\langle \sum_j \delta(z-z_j)\Bigr\rangle^{-1}
\nonumber 
\\
&=&
\frac{1}{\pi\rho(z)}
\lim_{\varepsilon\to 0}\left\langle \left(\tr
\frac{\varepsilon}{(z-{\cal H})(z^*-{\cal H}^\dagger)+\varepsilon^2}
\right)^2\right\rangle
\ ,
\end{eqnarray}
where 
$z=\Omega-i\Gamma/2$.
The eigenvalue density $\rho(z)=
\langle\sum_j \delta(z-z_j)\rangle$
has been determined 
by Fyodorov and Sommers \cite{fyodorov:97a},
\begin{eqnarray}
\label{Rhoexp}
&&\rho(z)=(2\pi/\Delta^2)
{\cal F}_1(\pi\Gamma/\Delta)\,{\cal F}_2(\pi\Gamma/\Delta)
\ ,
\\
\label{Fdef}
&&{\cal F}_1(y)=\frac{1}{(N-1)!}\,y^{N-1}\,e^{-gy}\quad,\quad
{\cal F}_2(y)=(-1)^N e^{gy}\, \left(\frac{d}{dy}\right)^N
\left(e^{-gy}\frac{\sinh y}{y}\right)\ .
\end{eqnarray}
The parameter $g\ge 1$ is related to the transmission probability $T$
of a scattering channel by $g=2/T-1$.\footnote{We assume for simplicity
that each channel has the same $g$. 
Expressions for ${\cal F}_{1,2}$
for channel-dependent $g_n$ can be found in ref.\ \cite{fyodorov:97a},
and our main result (\ref{Petfak5}) given below remains valid for the 
general case.}

The evaluation of the quantity (\ref{Petfak2}) proceeds by the 
supersymmetry approach. We introduce the generating functional 
$\langle\det(A-J)/\det(A+J)\rangle$, where $J$ is a source matrix and
\begin{equation}
\label{Adef}
A\equiv
\left(\begin{array}{cc}A_{11}&A_{12}\\
A_{21}&A_{22}\end{array}\right)=
\left(\begin{array}{cc}i\varepsilon & z-{\cal H} \\
z^*-{\cal H}^\dagger & i\varepsilon  \\
\end{array}\right)\ .
\end{equation}
The quantity $\pi\rho(z)\langle K\rangle=
-\langle \tr[(A^{-1})_{11}]\,\tr[(A^{-1})_{22}]\rangle$ 
can be expressed in terms of derivatives of the generating 
functional with respect to $J$. Following 
the technique of supersymmetry~\cite{efetov:97,fyodorov:97a}
one can express the 
Petermann factor as an integral over a $4\times 4$ matrix $Q$,
\begin{eqnarray}
\label{Petfak3}
\rho\,(z)\langle K\rangle 
&=&\frac{\pi}{4\Delta^2}\,\lim_{\varepsilon\to 0}
\,\int dQ\,e^{-{\cal L}(Q)}\,
\mbox{Str}\,(Q\sigma_x\,P_0\,P_+)\,\mbox{Str}\,(Q\sigma_x\,P_0\,P_-)
\ ,
\\
\label{Actdef}
{\cal L}(Q)&=&-\frac{\pi\varepsilon}{\Delta}\,\mbox{Str}\,(Q\sigma_x)-
\frac{\pi\Gamma}{2\Delta}\,\mbox{Str}\,(\sigma_z Q)+N
\mbox{Str}\,\ln(1+w Q \sigma_z)\ ,
\end{eqnarray}
where $w$ is related to $T$ by $T=4w(1+w)^{-2}$.
The matrix $Q$ obeys the non-linear constraint $Q^2=1$ 
and belongs to the coset space of the unitary 
non-linear $\sigma$-model~\cite{efetov:97}. 
It is a ``supermatrix'', meaning that it has an equal number of commuting
and anticommuting variables. The symbol 
``Str'' denotes the graded trace of a supermatrix,
$\mbox{Str}\,A=\mbox{tr}\,P_0 A$,
where $P_0=\mbox{diag}(1,-1,1,-1)$.
The Pauli-matrices $\sigma_x$, $\sigma_z$ are embedded in the 
$4\times 4$ supermatrix space as $\sigma_z=\mbox{diag}(1,1,-1,-1)$ and 
$(\sigma_x)_{jk}=\delta_{j,k+2}+\delta_{j+2,k}$. Furthermore,
$P_\pm=\frac{1}{2}(1\pm \sigma_z)$.

The integral (\ref{Petfak3}) can be evaluated by introducing Efetov's
parameterization of the matrix $Q$ for
the unitary $\sigma$-model \cite{efetov:97}.
The result is\footnote{For $\Gamma$ real positive
the integrals (\ref{Petfak3}) and (\ref{Petfak4}) are formally divergent.
The proper procedure for the further evaluation \cite{fyodorov:97a} is 
to choose $\Gamma$ imaginary and to perform the analytic 
continuation at the end of the calculation.}
\begin{eqnarray}
\label{Petfak4}
\rho\,(z)\langle K\rangle
&=&-\frac{\pi}{2\Delta^2}\lim_{\varepsilon\to 0}\,
\varepsilon\dif{}{\varepsilon}
\int_1^\infty d\lambda_1 \int_{-1}^1 
d\lambda_2\,\frac{\lambda_1+\lambda_2}{\lambda_1-\lambda_2}
\left(\frac{g+\lambda_2}{g+\lambda_1}\right)^N
\nonumber \\
&&{}\times 
J_0\left(\varepsilon\sqrt{\lambda_1^2-1}\right)
\exp\left[\frac{\pi\Gamma}{\Delta}(\lambda_1-\lambda_2)\right]
\ ,
\end{eqnarray}
where $J_0$ is a Bessel function.
The remaining two ordinary integrals can be carried out as in the
corresponding calculation of $\rho(z)$ \cite{fyodorov:97a}.
We finally obtain 
\begin{equation}
\label{Petfak5}
\langle K\rangle=1+\frac{2\, S(y)}
{{\cal F}_1(y){\cal F}_2(y)}\quad,\quad 
S(y)= -\int_0^y 
dx\ {\cal F}_1(x)\,\dif{}{x}\,{\cal F}_2(x)\quad,\quad
y=\pi\Gamma/\Delta\ .
\end{equation}
A more explicit expression can be obtained by successive 
partial integrations:
\begin{equation}
\label{Sres1}
S(y)=\sum_{k=0}^{N-1}\frac{(-1)^k}{k!}\,y^k\,\left(\dif{}{y}\right)^k
\,\left(e^{-gy}\,\dif{}{y}\,\frac{\sinh y}{y} \right)\ .
\end{equation}
For $N=1$ and $\Gamma\ll\Delta$ we recover the single-channel result
of ref.\ \cite{patra:99},
$\langle K\rangle=1+2\pi\Gamma/3g\Delta$.
Eq.\ (\ref{Petfak5}) generalizes this result to any value of $N$
and $\Gamma$.

To study the limit $N\gg 1$, 
it is convenient to derive integral expressions 
by replacing $\sinh y/y$ in ${\cal F}_2$ with 
$\frac{1}{2}\int_{-1}^1 d\lambda\,e^{-\lambda y}$. This was done
in ref.\ \cite{fyodorov:97a} for the eigenvalue density, with the
result
\begin{equation}
\label{Rhoint}
{\cal F}_1(y){\cal F}_2(y)
=\frac{1}{2\,y^2\,(N-1)!}\,\int_{(g-1)y}^{(g+1)y} 
dt\ t^N\,e^{-t}\ .
\end{equation}
Similarly, we find for the function $S(y)$ that determines the
Petermann factor the expression
\begin{equation}
\label{Sint}
S(y)=-\frac{1}{4\,y^2\,(N-1)!}\int_{(g-1)y}^{(g+1)y}
dt\ [t-(g-1)y][t-(g+1)y]\,t^{N-1}\,e^{-t}\ .
\end{equation}
For simplicity we restrict ourselves in what follows to the case $T=g=1$
of fully transmitted scattering channels. When
$N\gg 1$, the factor $t^N\,e^{-t}$ has a sharp maximum at 
$t= N$ and one can apply the saddle-point method to evaluate the 
$t$-integration. This gives the eigenvalue density
\cite{fyodorov:97a} $\rho\,(z)\approx N/\pi\Gamma^2$ for
$\Gamma>\Gamma_0$ (with $\Gamma_0=N\Delta/2\pi$ for $T=1$).
The density vanishes exponentially as $\Gamma$ drops
below $\Gamma_0$.
Application of the same saddle-point method to eq.\ (\ref{Sres1})
gives $\langle K\rangle\approx(2\pi/\Delta)(\Gamma-\Gamma_0)$.
If $\Gamma-\Gamma_0=\order(\Gamma_0)$, then this estimate would imply
that $\langle K\rangle\propto N$. This linear scaling with $N$ breaks down
in the cutoff regime $\Gamma\lesssim\Gamma_0$, 
which is precisely the relevant regime for the laser cavities.

In order to study the cutoff regime we define the rescaled decay rate 
$u=\sqrt{N/2}(\Gamma/\Gamma_0-1)$ and 
take the limit $N\gg 1$ at fixed $u$. Expanding the integrands in eqs.\
(\ref{Rhoint}) and (\ref{Sint}) around the saddle-point and keeping the first 
non-Gaussian correction, we obtain
\begin{equation}
\label{Petfak8}
\langle K\rangle =\sqrt{2N}[F(u)+u]+
F(u)\left[2u+{\textstyle \frac{4}{3}}u^3+
{\textstyle \frac{4}{3}}(1+u^2)\,F(u)\right]
+
\order(N^{-1/2})
\ ,
\end{equation}
where we have abbreviated $F(u)= \pi^{-1/2}\exp[-u^2]
[1+\mbox{erf}\,(u)]^{-1}$.
The eigenvalue density in the cutoff regime has the form
$ \rho= (2\pi/N\Delta^2)[1+\mbox{erf}\,(u)]+\order(N^{-3/2})$.
For $u=0$ (hence $\Gamma=\Gamma_0$) eq.\ (\ref{Petfak8}) simplifies to
\begin{equation}
\langle K\rangle_{u=0}=\sqrt{\frac{2N}{\pi}}+\frac{4}{3\pi}\ ,
\label{Petfak10}
\end{equation}
while for $u\ll -1$ (hence $\Gamma-\Gamma_0\ll-\sqrt{N} \Delta$) we find
\begin{equation}
\langle K\rangle_{u\ll -1}=-\frac{1}{u}\sqrt{\frac{N}{2}}
=\frac{\Gamma_0}{\Gamma_0-\Gamma}\ .
\end{equation}

To determine how $\langle K\rangle$ scales with $N$ we need to estimate
the most probable value of $\Gamma$ for the lasing mode. The decay rate
$\Gamma_l$ of the lasing mode is the smallest among the cavity modes
that are amplified by stimulated emission. The number $m_a$ of
amplified modes is $\gg N$ in the ``good cavity''
regime\footnote{A ``good cavity'' has a smallest typical decay rate
(inverse mean dwell time) $\Gamma_0$ that is small compared to the
amplification bandwidth $\Omega_a$. Since $m_a\simeq\Omega_a/\Delta$
and $\Gamma_0=N\Delta/2\pi$, it follows that $m_a\gg N$.}.
In a simplified model \cite{misirpashaev:98a}, we can determine 
$\Gamma_l$ as  
the smallest of $m_a$
independent random variables $\Gamma_1,\ldots,\Gamma_{m_a}$
with probability
distribution
$P(\Gamma)=(\Delta/2)\rho(z)\approx(1/2\Gamma_0)[1+\mbox{erf}\,(u)]$.
The distribution of $\Gamma_l$ is then given by
\begin{equation}
\label{Mindist}
P_l(\Gamma_l)=
m_a\,P(\Gamma_l)\,
\left[1-\int_0^{\Gamma_l} d\Gamma\,P(\Gamma)\right]^{m_a-1}.
\end{equation}
This distribution has a pronounced maximum at a value $u_{\rm max}$ 
determined by 
\begin{equation}
\label{Maxeq}
\frac{\exp[-u_{\rm max}^2]}{[1+\mbox{erf}(u_{\rm max})]^2}=
\frac{m_a-1}{2}\sqrt{\frac{\pi}{2N}}\ .
\end{equation}
If $m_a$ is comparable to $\sqrt{N}$, we have 
$u_{\rm max}=\order(1)$ and the Petermann factor 
is to close to its value at $u=0$, eq.\ (\ref{Petfak10}).
For $m_a\gg \sqrt{N}$ (which is realized in the good-cavity regime)
we find 
$u_{\rm max}\sim
-\sqrt{\ln(m_a/\sqrt{N})}\sim -\sqrt{\ln m_a}\ll-1$ and 
\begin{equation}
\label{Petfak11}
\langle K\rangle\sim \sqrt{\frac{N}{\ln m_a}}
\ .
\end{equation}
This is still parametrically larger than unity as long as $m_a\ll e^N$. 

\begin{figure}
\epsfig{file=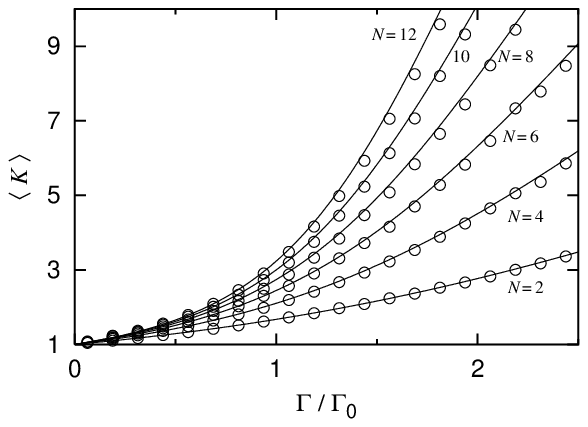,width=12cm}\\
\begin{picture}(1,1)
\put(51,130){\epsfig{file=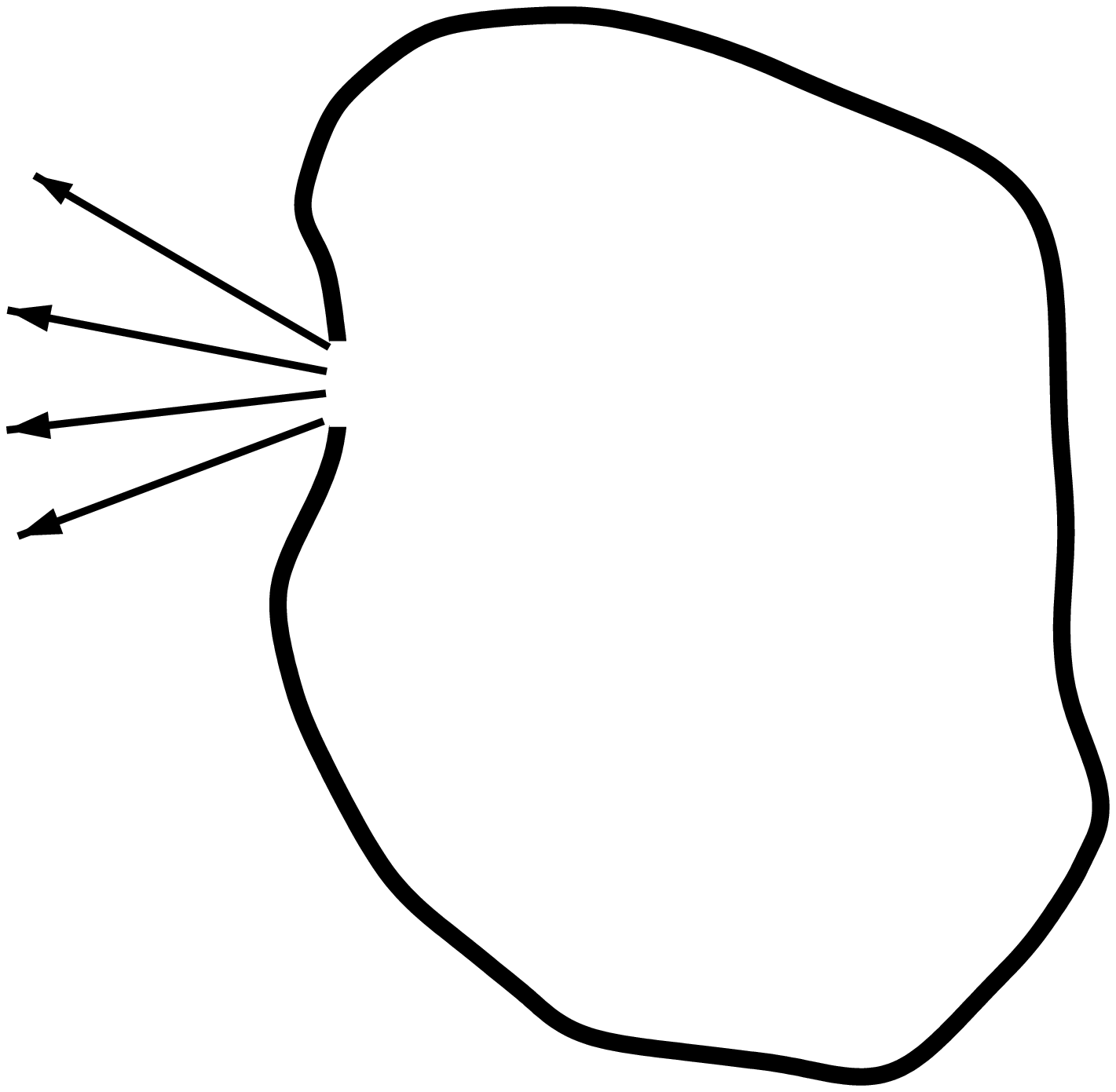,width=3cm}}
\end{picture}
\caption{Average Petermann factor $\langle K\rangle$ as a function of the
decay rate $\Gamma$ (in units of $\Gamma_0=N\Delta/2\pi$)
for a chaotic cavity (inset) having an
opening which supports $N$ fully transmitted scattering channels.
The solid curves are the analytical result~(\protect\ref{Petfak5}),
the data points are a numerical simulation.
}
\label{fig:mean1}
\end{figure}

\begin{figure}
\epsfig{file=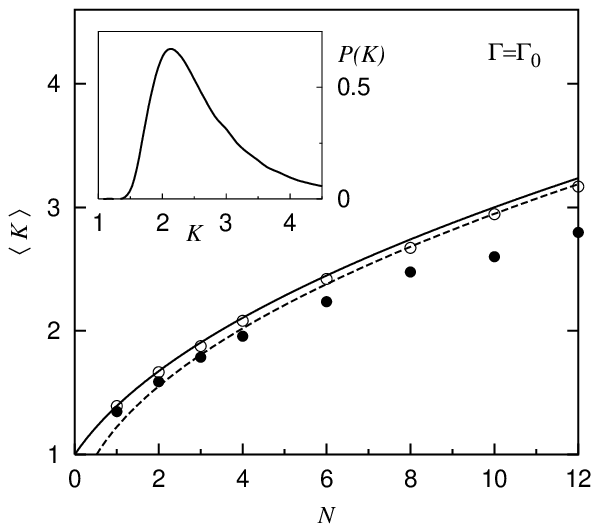,width=12cm}
\caption{Average Petermann factor at $\Gamma=\Gamma_0$ as
a function of $N$. The exact 
result (\protect\ref{Petfak5}) for broken time-reversal symmetry (solid curve)
is compared with the result of a numerical
simulation (open circles). Also shown is the simplified expression 
(\protect\ref{Petfak10}) for 
large $N$ (dashed curve). The solid circles are the numerical
result for preserved time-reversal symmetry. 
The inset shows the distribution of
$K$ at $\Gamma=\Gamma_0$ for $N=10$ and broken time-reversal symmetry,
following from the numerical simulation. 
}
\label{fig:ksqrtn}
\end{figure}

We now compare our analytical findings with
the results of numerical simulations.
We generated a large number of random matrices ${\cal H}$ with
dimension $M=120$ ($M=200$) for $N=2,4,6,8$ ($N=10,12$).
Figure \ref{fig:mean1} shows the average of $K$ at given $\Gamma$.
We find excellent
agreement with our analytical result (\ref{Petfak5}).
The scaling $\langle K\rangle\propto \sqrt{N}$ at $\Gamma=\Gamma_0$
is shown in fig.\ \ref{fig:ksqrtn}.
The inset depicts the
distribution of $K$ at $\Gamma=\Gamma_0$ for $N=10$,
which can only be accessed numerically. We see that the mean Petermann
factor is somewhat larger than the most probable value.

So far we only discussed the case of broken time-reversal symmetry,
because this greatly simplifies the analysis.
From a numerical simulation
with real symmetric $H$ (shown as well in fig.\ \ref{fig:ksqrtn})
we find that the average Petermann factor
for $\Gamma=\Gamma_0$ is parametrically large
also in the presence of time-reversal symmetry.
The average increases again
sublinearly with $N$, but with an exponent that is smaller than in the
case of broken time-reversal symmetry. 

In summary, we have calculated the average Petermann factor $\langle
K\rangle$
for a chaotic cavity in the case of broken time-reversal symmetry. We
find that for typical values of the decay rate $\Gamma$ the average
$\langle K\rangle\propto\sqrt{N}$ scales with the square root of the
number of scattering channels $N$, hence with the square root of the area
${\cal A}$ of the opening of the cavity.
This result could only be obtained by a non-perturbative calculation
because laser action selects a mode with an untypically small decay rate.
From a numerical simulation we find that the sublinear increase
holds also in the case of preserved time-reversal symmetry.
The quantity $K$ deserves interest not only because it determines the
quantum limit of the linewidth of laser radiation, via eq.\ (\ref{eq:k}).
It is of fundamental significance in the general
theory of scattering resonances, where it enters the width-to-height
relation
of resonance peaks 
and determines the scattering strength of a quasi-bound state with given
decay
rate $\Gamma$, via eq.\ (\ref{eq:sr2}).
Because $K$ is a measure for the non-orthogonality of the 
eigenmodes, this provides a way to obtain additional information 
about an externally probed open system. 

This work was supported by the Dutch Science Foundation NWO/FOM and by the
TMR program of the European Commission. The authors thank the Max 
Planck Institute for Physics of Complex Systems in Dresden for its hospitality 
during the workshop: {\em Dynamics of Complex Systems} at which part of
this work was done.

\end{document}